%% file: main.tex
\documentclass[conference]{IEEEtran}
\IEEEoverridecommandlockouts
\usepackage{cite}
\usepackage{amsmath,amssymb,amsfonts}
\usepackage{algorithmic}
\usepackage{graphicx}
\usepackage{textcomp}
\usepackage{xcolor}
\usepackage{float}
\usepackage{booktabs}

\usepackage{acronym}
\usepackage{subcaption}
\usepackage{float}
\usepackage{multirow}
\usepackage{soul}
\usepackage{comment}
\usepackage{url}

\usepackage{enumitem,amssymb}
\newlist{todolist}{itemize}{2}
\setlist[todolist]{label=$\square$}
\usepackage{pifont}

\usepackage{array}
\newcolumntype{P}[1]{>{\centering\arraybackslash}p{#1}}
\newcolumntype{M}[1]{>{\centering\arraybackslash}m{#1}}

\def\BibTeX{{\rm B\kern-.05em{\sc i\kern-.025em b}\kern-.08em
    T\kern-.1667em\lower.7ex\hbox{E}\kern-.125emX}}
\begin{document}

\acrodef{FR}{frame rate}
\acrodef{WULPUS}{Wearable Ultra Low-Power Ultrasound}
\acrodef{US}{Ultrasound}
\acrodef{ULP}{ultra low-power}
\acrodef{AFE}{analog front-end}
\acrodef{RF}{radio frequency}
\acrodef{ML}{machine Learning}
\acrodef{FPGA}{Field-Programmable Gate Array}
\acrodef{LVDS}{low-voltage differential signalling}
\acrodef{SPI}{serial peripheral interface}
\acrodef{MCU}{microcontroller unit}
\acrodef{ADC}{analog to digital Converter}
\acrodef{HDL}{hardware description language}
\acrodef{DMA}{direct memory access}

\acrodef{HD}{hardware design}
\acrodef{AD}{algorithm development}
\acrodef{FPS}{frames per second}
\acrodef{DAS}{delay-and-sum}
\acrodef{TGC}{time-gain compensation}

\acrodef{PULP}{Parallel Ultra Low Power}
\acrodef{SoC}{System on Chip}
\acrodef{BLE}{Bluetooth Low Energy}
\acrodef{EEG}{electroencephalography}
\acrodef{EMG}{electromyography}
\acrodef{ECG}{electrocardiogram}

\title{Open-Source Live-Reconfigurable Multi-Mode Wearable Ultrasound
\thanks{We acknowledge support from the ETH Research Grant ETH-C-01-21-2 (Project ListenToLight) and from the ETH Future Computing Lab (EFCL).}
}

\author{
  \IEEEauthorblockN{
    Cédric~Hirschi\IEEEauthorrefmark{1},
    Federico~Villani\IEEEauthorrefmark{1},
    Luca~Benini\IEEEauthorrefmark{1}\IEEEauthorrefmark{2},
    Andrea~Cossettini\IEEEauthorrefmark{1}\\}

  \IEEEauthorblockA{\IEEEauthorrefmark{1}Integrated Systems Laboratory, ETH Z{\"u}rich, Z{\"u}rich, Switzerland}
  \IEEEauthorblockA{\IEEEauthorrefmark{2}DEI, University of Bologna, Bologna, Italy}
}

\maketitle

\begin{abstract}

Wearable ultrasound enables continuous deep-tissue monitoring, and a single programmable probe can operate in multiple complementary modes, such as structural A-mode and Doppler flow measurement. 
However, each operating mode requires dedicated measurement parameters and peripheral states, with no single configuration serving all modes on resource-constrained devices. Time multiplexing of operating modes introduces reconfiguration latency that lowers the effective mode repetition rate. 
To address this limitation, we present an open-source, transition-aware control stack for low-latency, in-session reconfiguration of the 32-channel TinyProbe wearable platform.
Operating modes are described as hardware configurations, and host-side shadow registers track the peripheral states, enabling transition-specific register updates. 
Transition sequences are executed either by the host (over Wi-Fi 6) or by a firmware loop on the probe MCU.
We validate the stack on a pulsatile-flow phantom by interleaving blocks of 25 to 100 pulsed-wave Doppler shots at 1.43 kHz PRF with single 16-channel A-mode acquisitions, changing channel configurations at every transition. 
Compared to full reconfiguration, the overhead per transition decreases from 30.2 ms to 11.6 ms (host-scheduled) and 3.1 ms (MCU-scheduled).
For 75-shot Doppler blocks, the multi-mode repetition rate reaches 16.0 Hz (MCU-scheduled), 90.1\% of the theoretical maximum of 17.7 Hz. Concurrent reconstruction of a Doppler spectrogram and a lumen-diameter trace demonstrates the functionality of time-multiplexed flow and structural monitoring.

\end{abstract}


\input{sections/intro}
\input{sections/methods}
\input{sections/resultsdiscussion}
\input{sections/conclusion}

\section*{Acknowledgment}

We thank Thomas Quanbrough (ETH Z{\"u}rich) for technical support.

\bibliographystyle{IEEEtran}
\bibliography{bibliography.bib}

\end{document}

%% file: sections/intro.tex
\section{Introduction}

Wearable ultrasound (WUS) enables continuous, non-invasive monitoring of physiological processes beneath the skin~\cite{zhou2025wearable}. Unlike wearable sensing modalities like photoplethysmography and biopotential sensing, ultrasound probes deep tissue. A single programmable wearable probe can operate in A-mode, M-mode, B-mode imaging, and Doppler flow measurements, enabling applications across cardiovascular monitoring, muscle activity sensing, and human-machine interfaces~\cite{kenny2021handsfree, yang2021kinematics, sgambato2024hmi,gao2024wearable,hu2023wearable,lu2026hand}.

Most WUS systems remain application-specific and operate with fixed or narrowly configurable acquisition pipelines~\cite{frey2022wulpus,kenny2021handsfree,lin2024integrated}. Recent platforms such as TinyProbe~\cite{vostrikov2025tinyprobe} expose multiple software-defined operating modes with on-host hardware representation, but published demonstrations configure and evaluate each mode separately rather than switching between modes within one acquisition session. With the existing control flow, each mode switch requires the host to rewrite all of the FPGA, analog front end (AFE), and pulser settings before acquisition resumes, adding latency at each transition.

On resource-constrained wearables, different operating modes generally cannot share one configuration. An example application requiring such switching is the joint assessment of vascular structure and blood flow. In fact, unlike cabled research scanners, a WUS probe must adapt the number of active channels, samples per shot, and PRF to the limited wireless throughput~\cite{hirschi2025wifi6}: pulsed-wave Doppler requires a \mbox{kHz-range} PRF to measure arterial blood flow and can therefore stream only a few receive channels, whereas structural imaging benefits from higher channel counts to resolve vessel geometry and can therefore operate only at a lower frame rate~\cite{vostrikov2025tinyprobe}. 
While time multiplexing offers a practical approach by interleaving acquisition modes within one session, switching introduces reconfiguration latency, reduces the effective repetition rate, and introduces gaps in the physiological signal.

In this work, we present an open-source, transition-aware control stack for in-session reconfiguration of TinyProbe. The main contributions are:

\begin{itemize}
    \item A state-aware hardware abstraction that converts complete operating mode configurations into transition-specific peripheral operations.
    \item A multi-mode schedule planner supporting block-based acquisition, hardware-profile reuse, and cyclic sequence caching.
    \item Experimental validation showing a reduction in mean transition overhead from 30.2 ms to 11.6 ms (host-scheduled) and 3.1 ms (MCU-scheduled), increasing the repetition rate of interleaved acquisitions (75 Doppler shots at 1.43 kHz PRF and a single structural frame) from 8.6 Hz to 12.5 Hz and 16.0 Hz, respectively.
    \item Open-source release of the control stack at \url{https://github.com/pulp-bio/TinyProbe}
\end{itemize}

%% file: sections/methods.tex
\section{Materials and Methods}

\subsection{TinyProbe Architecture}

The proposed control stack is implemented on the \mbox{32-channel} TinyProbe platform~\cite{vostrikov2025tinyprobe}, using the Wi-Fi 6 system controller introduced in~\cite{hirschi2025wifi6}. The platform combines a wearable form factor for continuous, on-body operation with a runtime-programmable acquisition pipeline. The components relevant to runtime reconfiguration are the TX7332 pulser, AFE5832LP analog front end, FPGA acquisition subsystem, and SiWG917 Wi-Fi 6 microcontroller (MCU).

\begin{figure}[htb]
    \centering
    \includegraphics[width=0.4\textwidth]{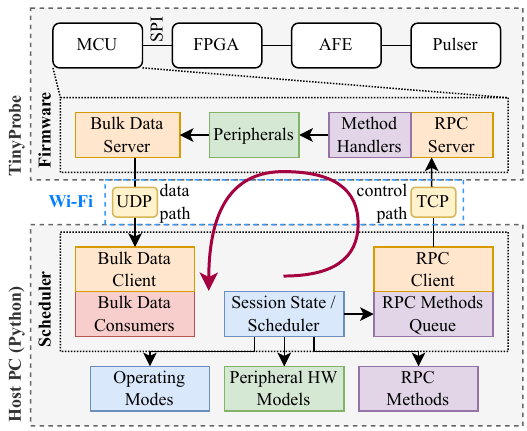}
    \caption{Control-stack architecture. Host-side mode configurations generate RPCs sent over TCP to the MCU, which configures peripherals via SPI. Acquired data from the AFE returns to the host via FPGA and MCU over UDP.}
    \label{fig:system_overview}
\end{figure}

Fig.~\ref{fig:system_overview} shows the system architecture of the control on the host and the MCU. In a measurement, the system goes through configuration and acquisition phases. During configuration, the scheduler on the host batches the configuration for the FPGA, AFE, and pulser (peripherals) into remote procedure call (RPC) methods, which are sent via the TCP control path over Wi-Fi to the MCU firmware. The MCU firmware handles RPCs by executing each method, checking whether it succeeded, and returning all statuses to the waiting host after execution. These methods include SPI-target selection, FPGA, AFE, and pulser register writes, power control, delays, and acquisition triggering.

The peripherals are all connected via SPI, with a MUX integrated on the FPGA, routing requests to each peripheral as controlled by the MCU.

Once an acquisition is started, the FPGA handles the acquisition events and timing, acquires data from the AFE via LVDS, buffers the data, and relays it to the MCU via SPI. There, the data is served as packets for the UDP-based bulk-data path, which sustains throughputs of up to 35.7 Mbps~\cite{hirschi2025wifi6}.

\subsection{State-Aware Hardware Control}

The lowest host-side layer provides peripheral hardware models as abstraction layers for the FPGA, AFE, and pulser. These layers expose device parameters as typed parameters and map them onto the corresponding register fields. Setting a parameter updates a host-side shadow register representation instead of immediately communicating with the probe.

When a field in a shadow register map is written via a shared session, its containing register is marked as dirty. These changes are batched into RPC methods at the end of the configuration phase, and modified fields belonging to the same register are combined into a single value. This coalesces all parameter changes into the minimum number of register writes, which the MCU executes. The corresponding dirty flag is cleared after the MCU returns a successful method status.

The session tracks the currently selected SPI target and inserts a multiplexer-switch operation only when the target changes. All register writes are grouped by peripheral to reduce the number of switches. Under the assumption that all hardware modifications are performed through the shared session, the register shadows provide a host-side view of the hardware state and avoid complete register readback between acquisition phases. To mitigate the risk of divergence between the shadow state and the actual hardware state, a register write can optionally include an immediate readback to verify that the write was successful. This adds overhead and is therefore disabled during the final measurements.

\subsection{Operating Mode and Transition Planning}

We define an operating \textit{mode} as the complete set of FPGA, AFE, pulser, and data-transfer parameters required for a particular US measurement. An \textit{acquisition block} contains one or more consecutive shots performed in one operating mode. A \textit{schedule} is an ordered sequence of acquisition blocks, and a \textit{transition} comprises the control operations required to change from one operating mode to the next.

A three-phase flow (setup, acquisition, and teardown) is defined in Python using these parameters. Planning a phase applies a set of parameters to the shared hardware models and collects the resulting register updates and control operations without immediately transmitting them to the probe. The planned phase can then be passed to an execution backend.

During setup, the first block of a schedule is fully configured, and the pulser delay and pattern profiles used by all modes are preloaded into the pulser profile memories. After the planner computes all transitions, the resulting method sequences are cached and reused for subsequent transitions, avoiding repeated host-side planning.

A transition places the affected subsystems in a state where they accept configuration changes, applies only the required register or profile changes, and restores operation. Transitions between pulser configurations select a preloaded delay and pattern profile instead of rewriting the associated profile memory.

\subsection{Execution Backends}

\subsubsection{Host-Scheduled}

The host-scheduled backend executes the planned method sequences exactly as they are emitted. Long RPC requests are partitioned according to the maximum packet size allowed by the network link.
Consequently, TCP connection setup, packet transmission, and response latency contribute to the switching overhead (11.6 ms, see below), whereas the raw-data throughput (35.7 Mbps) is determined by the independent UDP path.

\subsubsection{MCU-Scheduled}

For the MCU-scheduled backend, the host can mark a section that should be looped, for example, the complete block sequence. For this, a command is inserted into the method queue with a given repetition count and a target index to loop back to in the method queue. This way, the control communication between blocks is completely removed, reducing the switching overhead during acquisitions.

\subsection{Experimental Validation}

\begin{figure}[htb]
    \centering
    \includegraphics[width=\columnwidth]{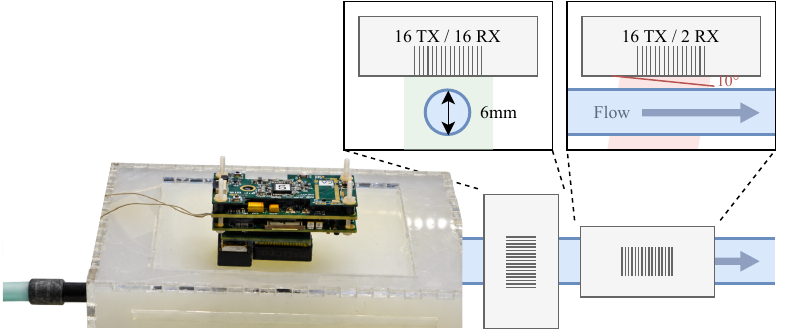}
    \caption{Setup (left) and arrangement of the two transducers relative to the observed artery phantom. The perpendicular transducer emits a 0° plane wave on 16 channels, and receives on 16 channels. The parallel transducer emits a 10° plane wave on 16 channels, and receives on 2 central channels.}
    \label{fig:experimental_setup}
\end{figure}

\textbf{Setup.}
We validate the live reconfiguration by placing TinyProbe, equipped with two 32-channel, 2.25 MHz linear-array transducers (LA-2.25-32, Vermon), on a custom artery phantom coupled with the DOPFLOW pulsatile flow system~\cite{dopflow} (see Fig.~\ref{fig:experimental_setup}). The phantom is cast from tissue-mimicking silicone rubber (EcoFlex 30, Smooth-On, USA) and contains a round, wall-less channel with a diameter of 6 mm, positioned 12 mm beneath the phantom surface, filled with a blood-mimicking fluid (GAMPT VK-50140). A syringe pump generates pulsatile flow at 60 BPM with a peak velocity of 0.7 m/s, pushing the fluid through the phantom into an open reservoir.

The two transducers are arranged in a T-shaped configuration, providing mode-specific views (Fig.~\ref{fig:experimental_setup}): the longitudinal transducer is aligned with the vessel for Doppler velocity estimation, whereas the transverse transducer is dedicated to structural diameter measurement. 
The central 16 elements of each transducer are connected to separate subsets of the 32 TinyProbe channels, so that each transition also switches the active transmit and receive channel sets. Fig.~\ref{fig:experimental_setup} summarizes the TX/RX configuration of each transducer.

\textbf{Transition Latency Measurement.}
The schedule repeats a cycle of one Doppler block of $N_D=\{25,50,75,100\}$ shots at a pulse repetition frequency of $\mathrm{PRF}_D=1430~\mathrm{Hz}$, followed by one structural acquisition lasting $T_S=4~\mathrm{ms}$. Without reconfiguration overhead, the cycle duration and the
corresponding maximum cycle rate are:
$$ T_{\mathrm{theo}} = \frac{N_D}{\mathrm{PRF}_D} + T_S, \hspace{1.5em} \mathrm{RR}_{\mathrm{theo}} = \frac{1}{T_{\mathrm{theo}}} = \frac{1}{\frac{N_D}{f_D} + \frac{1}{f_S}} $$
We measure the block repetition rate from MCU timestamps recorded at the start of each Doppler block. For each backend and each $N_D$, eight consecutive frames are recorded, yielding seven complete cycle intervals. The mean measured cycle duration $\overline{T}_{\mathrm{meas}}$ is computed from these intervals, and the corresponding repetition rate is $\mathrm{RR}_{\mathrm{meas}}=1/\overline{T}_{\mathrm{meas}}$. Since each cycle contains two operating-mode transitions, Doppler-to-structural and structural-to-Doppler, the mean end-to-end overhead per transition is defined as
$$ \overline{t}_{\mathrm{trans}} = \frac{\overline{T}_{\mathrm{meas}}-T_{\mathrm{theo}}}{2}. $$
This overhead includes all excess time on the critical path between acquisitions, including device-side reconfiguration and, for host-scheduled execution, host-to-MCU communication. We perform these measurements by comparing three backends: full reconfiguration, host-scheduled execution, and MCU-scheduled execution.

\textbf{Signal Reconstruction.}
The Doppler and structural recordings are reconstructed offline from the acquired RF samples. For Doppler processing, each RF trace is converted to its analytic representation using a Hilbert transform and demodulated at the transmit center frequency. Stationary clutter is suppressed independently for each receive channel by subtracting the slow-time mean and removing the leading singular component~\cite{Udesen2006,Demene2015}. Candidate receive-channel and depth gates are ranked according to their directional Doppler energy relative to near-zero frequency clutter and their signal power. The samples within the highest-ranked gate are averaged to form a complex slow-time trace, which is high-pass filtered at 20 Hz. A two-sided short-time Fourier transform is then computed using 64-shot Hann windows with 75\% overlap. The Doppler frequency ($f_\mathrm{d}$) is converted to flow velocity according to $v = \frac{f_\mathrm{d}c}{2f_0\cos\theta}$, where $c$ is the assumed speed of sound and $\theta$ is the angle between the ultrasound beam and the flow direction.

For structural processing, each RF trace is converted to its analytic representation using a Hilbert transform and demodulated at the transmit center frequency. Near and far wall echoes are initialized from the temporal median of each receive channel, using the phantom's nominal 6 mm diameter to select plausible echo pairs. Their positions are then tracked in each frame by normalized template correlation with sub-sample interpolation. Channels with low correlation or implausible diameters are excluded, and the remaining diameter estimates are combined using a median and low-pass filtered at 3 Hz.

%% file: sections/resultsdiscussion.tex
\section{Results and Discussion}

\subsection{Transition Latency}

For $N_D=75$, host and MCU scheduling reduce mean transition overhead by 61.5\% and 89.8\%, respectively, relative to full reconfiguration (Table~\ref{tab:overheads}). MCU scheduling reaches 90.1\% of the overhead-free 17.7 Hz repetition rate and shows the lowest transition variability, consistent with eliminating variable latency from inter-block TCP control exchanges.

\begin{table}[htb]
    \caption{Resulting mean end-to-end reconfiguration overheads $\overline{t}_\mathrm{trans}$ with their variance $s_\mathrm{trans}$ and maximum repetition rates $\mathrm{RR}_\mathrm{meas}$ for each backend with $N_D = 75$, calculated from seven complete cycle intervals.}
    \label{tab:overheads}
    \centering
    \begin{tabular}{lrr}
        \toprule
        Backend & $\overline{t}_\mathrm{trans} \pm s_\mathrm{trans}~\mathrm{[ms]}$ & $\mathrm{RR}_\mathrm{meas}\ \mathrm{[Hz]}$ \\
        \midrule
        Full Reconfiguration & $30.20\pm8.54$ & $8.55\pm1.25$ \\
        Host-Scheduled       & $11.62\pm7.32$ & $12.54\pm2.30$ \\
        MCU-Scheduled        & $3.08\pm0.02$ & $15.96\pm0.01$ \\
        \bottomrule
    \end{tabular}
\end{table}

\begin{figure}[htb]
    \centering
    \includegraphics[width=0.8\columnwidth]{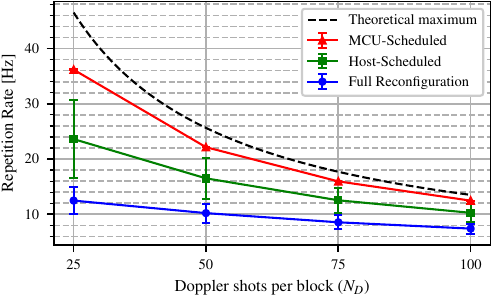}
    \caption{Mean measured repetition rates across all three configurations by amount of Doppler shots per frame calculated from 7 complete cycle intervals. Theoretical maximum $\mathrm{RR}_{\mathrm{theo}}$ shown for comparison. Error bars indicate standard deviation.}
    \label{fig:framerates}
\end{figure}

Fig.~\ref{fig:framerates} shows the achievable repetition rates for varying numbers of Doppler shots per frame. Since the transition overhead is approximately independent of block length, its relative contribution decreases with increasing $N_D$.

\subsection{Multimodal Acquisition}

\begin{figure}[htb]
    \centering
    \includegraphics[width=\columnwidth]{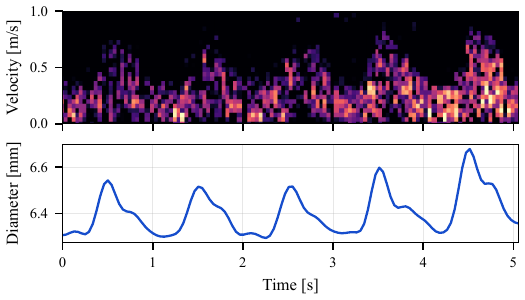}
    \caption{Simultaneously acquired doppler spectrogram (top) and lumen diameter trace (bottom) of the artery phantom in MCU-scheduled execution with $N_D = 75$. Simulated flow has a peak velocity of 0.7 m/s at 60 BPM.}
    \label{fig:multimodal}
\end{figure}

The resulting Doppler spectrogram and lumen-diameter trace are shown in Fig.~\ref{fig:multimodal}. Thanks to the low overhead of the MCU-scheduled execution used in this measurement, both quantities can be reconstructed without visible gaps in the displayed data despite time multiplexing the limited hardware resources.

%% file: sections/conclusion.tex
\section{Conclusion}

This paper presents an open-source, transition-aware control stack for enabling low-latency, in-session reconfiguration of multi-mode wearable ultrasound systems. Implemented on the 32-channel TinyProbe platform, the proposed architecture represents operating modes as complete hardware configurations and uses host-side device-state tracking to generate only the FPGA, AFE, and pulser reconfigurations required for each transition. Multi-mode acquisition schedules can be executed either from the host or autonomously by the probe MCU.

The proposed system reduces the mean end-to-end transition overhead from 30.2~ms with full reconfiguration to 11.6 ms for host-scheduled execution and 3.1 ms for MCU-scheduled execution, corresponding to reductions of 61.5\% and 89.8\%, respectively. For interleaved acquisitions consisting of 75 Doppler shots at 1.43 kHz PRF followed by one structural acquisition, this increases the repetition rate from 8.6 Hz to 12.5 Hz and 16.0 Hz, with MCU scheduling reaching 90.1\% of the theoretical maximum. The concurrent reconstruction of Doppler flow and lumen diameter further demonstrates that complementary functional and structural measurements can be time-multiplexed within a single acquisition session.

The remaining switching overhead is dominated by device-side reconfiguration and, for host-scheduled execution, network communication between acquisition blocks.

The control stack is publicly available at \url{https://github.com/pulp-bio/TinyProbe}.